\documentclass[10pt, notitlepage,nofootinbib,twocolumn]{revtex4-2}
\pdfoutput=1
\usepackage[left=1 in, right=1 in, top=1 in, bottom = 1 in]{geometry}
\usepackage[hidelinks]{hyperref}
\usepackage{graphicx}

\usepackage{amsmath,amssymb,mathtools,bm, physics}
\usepackage[colorinlistoftodos,prependcaption,textsize=tiny]{todonotes}
\usepackage{booktabs, hhline}
\usepackage{enumerate, enumitem}
\hypersetup{
    colorlinks=true,
    linkcolor=blue,
    filecolor=magenta,      	
    urlcolor=cyan,
}
\renewcommand{\r}{\vb r}

\begin{document}

\title{Adaptive Equilibration of Molecular Dynamics Simulations}

\author{Luciano G.~Silvestri}
  \email{silves28@msu.edu}
  \affiliation{Computational Mathematics, Science and Engineering, Michigan State University, East Lansing, Michigan 48824, USA}%
\author{Zachary A.~Johnson}
  \affiliation{Computational Mathematics, Science and Engineering, Michigan State University, East Lansing, Michigan 48824, USA}%
\author{Michael S.~Murillo}%
  \affiliation{Computational Mathematics, Science and Engineering, Michigan State University, East Lansing, Michigan 48824, USA}

\date{\today}

\begin{abstract}
We present a systematic framework for shortening and automating molecular dynamics equilibration through improved position initialization methods and uncertainty quantification analysis, using the Yukawa one-component plasma as an exemplar system. Our comprehensive evaluation of seven initialization approaches (uniform random, uniform random with rejection, Halton and Sobol sequences, perfect and perturbed lattices, and a Monte Carlo pair distribution method) demonstrates that initialization significantly impacts equilibration efficiency, with microfield distribution analysis providing diagnostic insights into thermal behaviors. Our results establish that initialization method selection is relatively inconsequential at low coupling strengths, while physics-informed methods demonstrate superior performance at high coupling strengths, reducing equilibration time. We establish direct relationships between temperature stability and uncertainties in transport properties (diffusion coefficient and viscosity), comparing thermostating protocols including ON-OFF versus OFF-ON duty cycles, Berendsen versus Langevin thermostats, and thermostat coupling strengths. Our findings demonstrate that weaker thermostat coupling generally requires fewer equilibration cycles, and OFF-ON thermostating sequences outperform ON-OFF approaches for most initialization methods. The methodology implements temperature forecasting as a quantitative metric for system thermalization, enabling users to determine equilibration adequacy based on specified uncertainty tolerances in desired output properties, thus transforming equilibration from a heuristic process to a rigorously quantifiable procedure with clear termination criteria.
\end{abstract}

\maketitle

\section{Introduction}
\label{sec:intro}
Molecular dynamics (MD) simulations constitute an important tool for investigating both classical and quantum many-body systems across physics, chemistry, materials science, and biology~\cite{AllenTildesley_Book_2017, Hollingsworth2018, 10.1063/1.5143225}. Obtaining physically meaningful results from these simulations requires an equilibration stage—a period during which the system reaches a stable, thermodynamically consistent state before data collection commences. This step is essential to ensure that the subsequent production run yields results that are neither biased by the initial configuration nor deviate from the target thermodynamic state~\cite{Braun2019}.

The efficiency of the equilibration phase is largely determined by the initial configuration of the system in phase space. In classical MD simulations without magnetic fields, including Born-Oppenheimer style density functional theory MD (DFT-MD) simulations, the phase space decouples and the velocity distribution is readily obtainable by sampling from the Maxwell-Boltzmann distribution. However, generating a set of initial positions consistent with the specified thermodynamic state presents a significantly greater challenge, necessitating an equilibration or thermalization phase where the system is driven to the required state, typically through the application of thermostats and/or barostats.

Traditionally, initial spatial configurations have been generated using methods appropriate for high-temperature situations such as uniform random placement (with or without a rejection radius), or methods that more closely approximate low-temperature configurations by starting from a known lattice. Such approaches are frequently accompanied by an initial optimization stage in the form of energy minimization. However, these methods can provide poor representations of the true desired state, resulting in characteristic temperature changes \cite{Kuzmin2002, gericke2003disorder}, and extended equilibration times or persistent biases in the resulting production runs. This has lead to recent work in improving the configuration initialization through running cheaper auxiliary MD simulations \cite{Fiedler2022}.

Despite the fundamental importance of the equilibration phase, the selection of equilibration parameters remains largely heuristic. Some knowledge of best practices exists \cite{Braun2019}, but researchers must often rely on experience, trial and error, or consultation with experts to determine appropriate thermostat strengths, equilibration durations, and thermostat algorithms. This lack of systematic methodology introduces potential inconsistencies across studies and raises questions about the reliability of subsequent production runs.

In this work, we tackle two main aspects of improving the equilibration of MD simulations. First, we reduce the arbitrary components of MD equilibration by casting the problem as a uncertainty quantification (UQ). In this framework, instead of guessing an adequate equilibration procedure \textit{a priori} we use estimates of the target of interest to inform how much equilibration is necessary. To illustrate our approach, we consider the example of the Yukawa one-component plasma for which the outputs of interest are transport coefficients such as self-diffusion and viscosity. We use approximate models for these numbers \cite{MURILLO200849, 10.1063/1.1316084} and use their numeric temperature dependence to turn MD temperature uncertainty into an uncertainty in the targeted output. This criterion-driven approach transforms equilibration from an open-ended preparatory step into a quantifiable part of the simulation with clear success/failure feedback.

The second main goal is a systematic look at initialization methodology, encompassing both established methods and novel techniques. We compare the amount of undesired heating and equilibration time for methods including uniform random, Monte Carlo-based techniques (uniform random with reject, and pair distribution random), low-discrepancy sequence methods, and lattice-based methods examining both perturbed and unperturbed lattices. Additionally, we compare the effectiveness of different thermostat algorithms, analyze the impact of thermostat coupling strengths, and explore cycling strategies (\textit{e.g.}, OFF-ON versus ON-OFF sequences).


The remainder of this paper is organized as follows: Section~\ref{sec:init_methods} provides comprehensive descriptions of the seven position initialization algorithms examined in this study along with their theoretical foundations and computational scaling properties. In section~\ref{sec:MD_methods} we introduce the Yukawa system as an example to investigate equilibration techniques, detail our simulation methods. We go into thermostating protocols, temperature deviation metrics and introduce an uncertainty quantification framework and statistical measures for establishing equilibration sufficiency based on target property uncertainties. Section~\ref{sec:results} presents a systematic comparative analysis of initialization-thermalization method combinations across multiple coupling regimes, including microfield distribution characterization, temperature forecasting validation, and quantitative assessment of thermostating duty cycles (ON-OFF versus OFF-ON). Finally, Section~\ref{sec:conclusions} synthesizes our findings into practical guidelines for optimizing MD equilibration.

\section{Position Initialization Methods} 
\label{sec:init_methods}

Molecular dynamics simulations are initialized with the specification of the positions and velocities of all particles. For classical simulations at equilibrium in the absence of a magnetic field, the phase space decouples and the specification of the velocities is simply that of drawing randomly from the Maxwell-Boltzmann distribution. In this section we describe several particle placement algorithms, summarized in Table~\ref{tab:initialization_methods}, with the goal of determining a set of best choices for certain physical situations.
\begin{table*}[t]
    \centering
    \begin{tabular}{|l|l|l|}
        \hline
        \textbf{Method} & \textbf{Description} & \textbf{Section} \\
        \hline
        Uni & Uniform random placement & \ref{subsec:random}  \\
        Uni Rej & Uniform random with rejection radius &  \ref{subsec:random_plus_reject} \\
        Halton & Low discrepancy, quasi-random sequence generator, Halton method &  \ref{subsubsec:Halton} \\
        Sobol & Low discrepancy, quasi-random sequence generator, Sobol method &  \ref{subsubsec:Sobol} \\
        MCPDF & Mesh based Monte Carlo method that matches input pair distribution function &  \ref{subsec:MCPDF}  \\
        BCC Uni & Body-centered cubic (BCC) lattice initialization &  \ref{subsec:lattice_initialization}  \\
        BCC Beta & BCC lattice with physical perturbations using compact beta function &  \ref{subsec:BCCBeta} \\
        \hline
    \end{tabular}
    \caption{Summary of initialization methods}
    \label{tab:initialization_methods}
\end{table*}
\subsection{Uniform Random}
\label{subsec:random}
One of the simplest methods for initializing the particles is to sample each coordinate uniformly from the available position space, $\mathbf r \sim \mathcal U(0, L_x)\mathcal U(0, L_y)\mathcal U(0, L_z)$. This approach is easy to implement, and the initialization itself is very fast and $\mathcal{O}(N)$. 

The drawbacks are a non-zero probability of coincident particle placement, which becomes increasingly likely with larger particle number. This clumping behavior can be quantified mathematically. For any two particles in a cubic simulation box with side length $L$, the probability that they fall within a distance $a$ of each other is approximately
\begin{equation}
P(d \leq a) \approx \frac{4\pi a^3}{3L^3}.
\end{equation}
While this probability may seem small for typical values of $a \ll L$, the scaling with particle number is problematic. For $N$ particles, we have $\binom{N}{2} = \frac{N(N-1)}{2}$ possible pairs, making the expected number of close pairs
\begin{equation}
E[\text{close pairs}] \approx \frac{2\pi a^3 N(N-1)}{3L^3},
\end{equation}
see Appendix~\ref{appendix:close_pair} for details. The quadratic scaling with $N$ means that as system size increases, clumping becomes virtually inevitable. The critical distance $a_c$ at which we expect to find the first close pair scales as $a_c \propto N^{-2/3}$, meaning that for large $N$, particles will be found at increasingly small separations. Such close placements result in large repulsive forces leading to substantial energy injection and subsequently long thermalization times.

\subsection{Uniform Random With Rejection}
\label{subsec:random_plus_reject}
The main issue with pure random placement is the possibility of coincident particles, which motivates a simple modification to the previous method, in which a rejection radius is added. If two particles are within a distance $r_{\rm rej}$ of each other, then the particle positions are resampled until their distance is $r > r_{\rm rej}$. 

The uniform random with rejection method directly addresses the clumping problem by enforcing a minimum separation between particles. From our analysis of clumping probability, an optimal choice of $r_{\text{rej}}$ would consider both the physical interaction potential and the scaling relationship $a_c \propto N^{-2/3}$, where $a_c$ is the critical distance at which clumping becomes likely. For systems with large $N$, this rejection method becomes increasingly necessary as the probability of finding at least one close pair approaches unity
\begin{equation}
P(\text{at least one close pair}) \approx 1 - e^{-\frac{2\pi a^3 N(N-1)}{3L^3}}
\end{equation}

While the algorithm for placing particles with a rejection radius may seem to be $\mathcal{O}(N^2)$ due to the pairwise distance checks, it can be made into $\mathcal{O}(N)$ with the use of linked cell lists.
\subsection{Low-Discrepancy Sequences}
\label{subsec:qmc_sequences}
Low-discrepancy sequences offer a powerful alternative for initializing particle positions and avoid clumping. These sequences, which include the Halton and Sobol sequences, are designed to uniformly cover the unit hypercube with minimal clustering, providing an efficient means of sampling that avoids the randomness of Monte Carlo methods. Low-discrepancy sequences are particularly advantageous in the high-temperature regime, where they can serve as an effective replacement for the uniform random with reject method. 

\subsubsection{Halton Sequence}
\label{subsubsec:Halton}

The Halton sequence \cite{halton_paper} is one of the simplest and most widely used low-discrepancy sequences. It is constructed by repeatedly dividing the unit interval according to prime bases, ensuring that the sequence fills the space uniformly without clustering. For a given dimension $d$, the Halton sequence $x_n$ is defined as
\begin{align}
x_n = \left( \phi_{b_1}(n), \phi_{b_2}(n), \ldots, \phi_{b_d}(n) \right),
\end{align}
where $\phi_b(n)$ is the radical inverse function for base $b$, given by
\begin{align}
\phi_b(n) = \sum_{k=0}^{\infty} a_k b^{-k-1}
\end{align}
Here, $n$ is the integer being transformed, and $a_k$ are the digits of $n$ in base $b$. The choice of different prime bases $b_1, b_2, \ldots, b_d$ ensures that the sequence covers the $d$-dimensional space uniformly [REF Needed].

\subsubsection{Sobol Sequence}
\label{subsubsec:Sobol}

The Sobol sequence \cite{sobol1967distribution} is another low-discrepancy sequence, known for its superior uniformity in higher dimensions compared to the Halton sequence. It is constructed using a set of direction numbers that define the sequence recursively, leading to an efficient filling of the unit hypercube. The Sobol sequence $x_n$ in $d$ dimensions is generated as
\begin{align}
x_n = \left( x_n^{(1)}, x_n^{(2)}, \ldots, x_n^{(d)} \right),
\end{align}
with each component $x_n^{(j)}$ given by
\begin{align}
x_n^{(j)} = \frac{n}{2} \oplus v_n^{(j)},
\end{align}
where $\oplus$ denotes bitwise \verb+XOR+, and $v_n^{(j)}$ are the direction numbers that depend on the dimension $j$. The recursive nature of the Sobol sequence ensures that each point added to the sequence improves the coverage of the space, making it particularly effective for high-dimensional integration tasks.

\subsection{Monte Carlo Matching of Pair Distribution Function}
\label{subsec:MCPDF}
The structure in the liquid to gas regime can be well approximated by the one and two-body correlations only. In homogeneous systems these are described by the density and the pair distribution function,
\begin{align}
g(\r,\r')  = \frac{\langle \sum_{i\neq j} \delta(\r_i - \r) \delta(\r_j - \r')\rangle}{n(\r)n(\r')},
\end{align} 
which, in the homogeneous limit become the average density, and the radial distribution function (RDF), $g(r) = g(|\r-\r'|)$. In this limit, the RDF approximates the $N$-body distribution function through the neglect of three-body correlations in a Kirkwood approximation \cite{Kirkwood1935},
\begin{align}
    g^{(N)}(\vb r_1, \cdots, \vb r_n) &\approx \prod_{\rm i>j} g^{(2)}(\vb r_i,\vb r_j) \\
       &\propto  \prod_{\rm i<n} g^{(2)}(\vb r_i,\vb r_N).
\end{align}
In the second line, we separate correlation contributions from the $N$-th position, which we imagine as the test location of a new probability. We will only need to consider the correlation of a single test point at $\r_N$ relative to all other particle positions. Thus, given the positions of $N-1$ particles, the probability distribution for placement of the $N$-th particle is
\begin{align}
\label{eq:probability_rdf}
    p(\r_N) &= \frac{g^{(N)}(\r_1,\cdots, \r_N)}{\int  d\r_N g^{(N)}(\r_1,\cdots, \r_N) }\\
            &\approx \frac{ \prod_{\rm i<N} g^{(2)}(\vb r_i,\vb r_N)}{ \int  d\r_N  \prod_{\rm i<N} g^{(2)}(\vb r_i,\vb r_N)}.
\end{align}

The key is that in many cases the pair distribution function is known to a good approximation by fits \cite{MORSALI200511, matteoli1995simple} or theory in the form of well studied approximations such as Hyper-Netted Chain (HNC) and Percus-Yevick (PY)\cite{hansen2013theory, ng1974hypernetted}. A best-guess RDF is then the only needed input. In this case, we use the HNC method in \cite{ng1974hypernetted} with a bridge function correction from \cite{daughton2000empirical}. 

One proceeds first by placing a single particle randomly in the domain. Each subsequent placements is made by drawing from a probability mesh defined using Eq.~\eqref{eq:probability_rdf}. Implementing this algorithm naively has a scaling of $N^2$, since for each particle we place, we must compute the distance of that particle to every gridpoint, the number of which scales as $N$. This can be simply averted, and order $N$ scaling achieved, if a correlation distance $r_{\rm corr}$ is used, chosen such that $g^{(2)}(r_{\rm corr})\approx 1$, and only considering mesh points within that distance using linked-cell lists or similar grouping methods.

For extremely large simulations, where one desires even sub-linear scaling, one can instead initialize only a subcell of $N_1$ particles, and then copy it to $M$ subcells per dimension and generate a configuration with $N=N_1 M^3$ total particles. This results in $\mathcal{N^0}$ scaling, as seen in Fig.~\ref{fig:methods_timing}. Periodic boundary conditions in the initial subcell prevent adjacent particles and the resulting distribution matches the input RDF exactly in the infinite particle, infinitely refined mesh limit. In order to avoid having identical subcells, we recommend adding small perturbations which also serves to smooth finite resolution RDFs. 

\subsection{Perfect Lattice Initialization}
\label{subsec:lattice_initialization}

Another widely used method is to initialize the particles on a regular lattice structure, such as a face-centered cubic (FCC) or body-centered cubic (BCC) lattice. This approach is particularly useful in systems where the particles are expected to exhibit crystalline order or when starting from a known solid-state structure. The lattice initialization ensures that particles are evenly spaced, avoiding the large force fluctuations that can arise from random placement.

Lattice initialization begins by placing particles in a lattice configuration that is consistent with the main simulation cell. The number of particles $N$ is selected to match the desired density $n$ for the chosen lattice type. For example, in a BCC lattice within a cubic simulation cell of volume $L^3$, the density is given by:
$$
n = \frac{N}{L^3}, \quad N = n_c \times 2
$$
where $n_c$ represents the number of unit cells along one edge. The positions of the particles are initially set to the lattice points $\mathbf{R}_i$. For the remainder of the paper this method will be identified as BCC Uni as we choose a BCC lattice. In order to ensure a non-zero force on the particles with add a very small random perturbation, sampled from a uniform distribution, to the particles positions.

\subsection{Perturbed Lattice Method}
\label{subsec:BCCBeta}

The perfect lattice method also has limitations. The ordered initial configuration may introduce artificial correlations that are not representative of the equilibrium state of a disordered system. As the system equilibrates, the lattice structure must relax, which can take considerable time depending on the nature of the interactions and the density of the system. Moreover, if the system is intended to simulate a fluid or amorphous state, the imposed initial order may bias the equilibration process, necessitating a thorough and extended equilibration period to allow the system to reach a truly disordered configuration. 

The perfect lattice method can be improved by incorporating thermal perturbations consistent with a given temperature. The perturbed lattice method (PLM) begins with the expectation value of a test particle density in the fixed potential of the remaining particles, assumed to be in a relevant lattice configuration. The test-particle density is given by
\begin{align}
\label{eqn:one_particle_density}
 \langle n({\bf r})\rangle  = {\cal C}\int d\r_i\: e^{-\beta U({\bf r}_i)/k_BT} \delta({\bf r}-{\bf r}_i),
\end{align}
where ${\cal C}$ is a normalization constant, and $U({\bf r}_i) = \sum_{j \neq i} u_{ij}({\bf r}_{ij}) + U_{\mathrm{ext}}({\bf r}_i)$ is the total potential felt by particle $i$ due to the remaining particles and a possible external potential. Each particle in the simulation is individually treated as a test particle with its position obtained by sampling $\langle n({\bf r})\rangle \sim P(\delta {\bf r})$. Note that the PLM is temperature aware, has the relevant lattice as the low-temperature limit, is interaction-potential aware, handles mixtures through $u_{ij}$, allows for an external potential, and is roughly consistent with the statistical mechanics of the particle density, while being an ${\cal O}(N)$ method. One can think of the PLM as an ``inverse radial distribution function'' in that the particles are placed at lattice sites and we ask where the particle near the origin is, rather than the reverse. 

In general, (\ref{eqn:one_particle_density}) is non-spherical and requires a numerical solution for a given $U({\bf r})$ and temperature $\beta^{-1}$. For particle placement reasons, however, we only require the total potential of particle $i$ near its lattice position ${\bf R}_i$; this allows us to perform an expansion. For small displacements $\mathbf{\delta r}_i$ around the equilibrium lattice points $\mathbf{R}_i$, the potential energy can be approximated by a harmonic potential. Expanding $U(\mathbf{r}_i)$ in a Taylor series up to second order, we obtain:
$$
U(\mathbf{r}_i) \approx U(\mathbf{R}_i) + \frac{1}{2} \mathbf{\delta r}_i^T \vb H_i \mathbf{\delta r}_i
$$
where $\mathbf{\delta r}_i = \mathbf{r}_i - \mathbf{R}_i$ and the curvature of the potential energy landscape is characterized by the Hessian matrix ${\bf H}_i$. This approximation is valid for small deviations from the lattice points, providing a quadratic potential well around each equilibrium position. Note that when this quadratic form is used in (\ref{eqn:one_particle_density}) the test-particle density is a multivariate Gaussian. The Hessian matrix ${\bf H}_i$ is explicitly given by
$$
\vb H_i = \sum_{j \neq i} \vb \nabla \otimes \vb \nabla u(|\mathbf{r}_i - \mathbf{r}_j|),
$$
where we ignore a possible external potential. Assuming local spherical symmetry (\textit{e.g.}, a cubic lattice) around each lattice site, the Hessian matrix $\vb H(u(-\mathbf{n}a))$ simplifies significantly. The Hessian for a spherically symmetric potential $u(r) = u(\|\mathbf{r}\|)$ in three dimensions is:
$$
\vb H(u(\mathbf{r})) = u''(r) \frac{\mathbf{r}\mathbf{r}^\top}{r^2} + u'(r) \left(\frac{\mathbf{I}}{r} - \frac{\mathbf{r}\mathbf{r}^\top}{r^3}\right)
$$
where $\mathbf{I}$ is the identity matrix and $r = \| \r \|$. This matrix accounts for both radial and tangential forces, offering a complete description of the local potential's curvature and the resulting forces on particles.

To model the effect of thermal fluctuations, we assume that particle displacements $\mathbf{\delta r}_i$ are distributed according to a Gaussian distribution. The variance of this distribution is determined by the temperature $T$ and the inverse of the Hessian matrix $\vb H_i$:
$$
P(\mathbf{\delta r}_i) \propto \exp\left(-\frac{1}{2k_B T} \mathbf{\delta r}_i^T \vb H_i \mathbf{\delta r}_i\right)
$$
The covariance matrix of this Gaussian distribution is given by:
$$
\vb C_i = k_B T \vb H_i^{-1}
$$
This allows for the generation of thermally perturbed initial positions for the particles by sampling from the Gaussian distribution.

In the high-temperature limit, this algorithm effectively reduces to nearly the uniform random method of subsection (\ref{subsec:random}), as the thermal perturbations dominate, causing particles to be uniformly sampled. Conversely, in the low-temperature limit, $T \sim 0$, the algorithm naturally converges to the perfect lattice configuration of subsection (\ref{subsec:lattice_initialization}) as thermal perturbations vanish, leaving particles in their equilibrium lattice positions. This approach smoothly transitions between these two limits, utilizing both the pair potential and the temperature to generate physically motivated initial configurations. 

\begin{figure}[ht]
    \centering
    \includegraphics[width=0.9\linewidth]{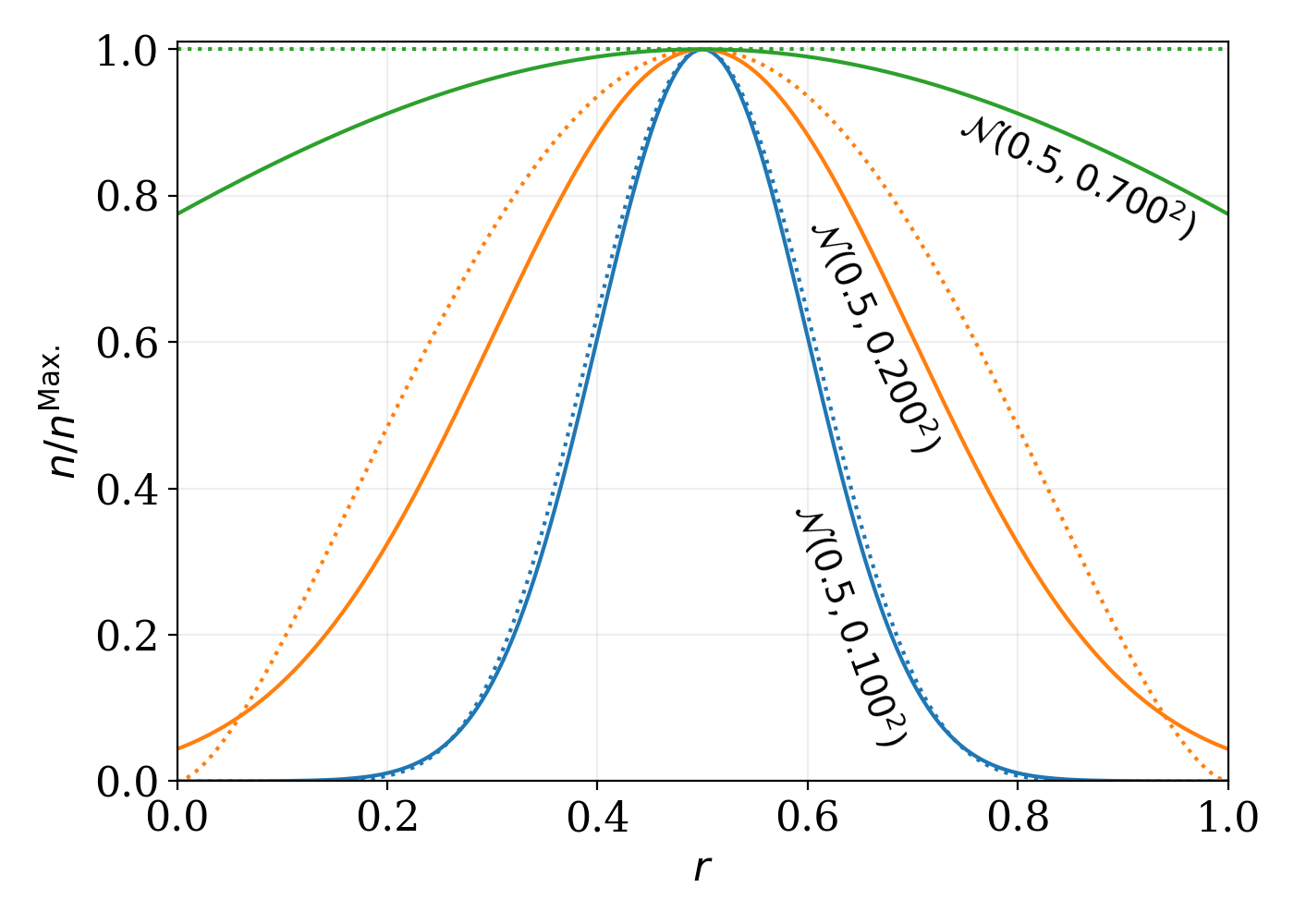}
    \caption{Comparison of the normal distribution (solid) and the compactly supported beta distribution (dotted) at three different standard deviation values. At low temperatures where the perturbations are small and the distribution has narrow width, the $\beta$ distribution approaches the normal distribution. }
    \label{fig:plm_comparison}
\end{figure}

Although it is straightforward to sample a multivariate Gaussian, such a protocol allows for the possibility that particles are placed very close to each other, especially at high temperature. To handle this limit smoothly, we employ the product of three (one for each cartesian coordinate) beta distributions with equal arguments, ${\mathcal Be} (\alpha_{x}, \alpha_{x}){\mathcal Be} (\alpha_{y}, \alpha_{y}){\mathcal Be} (\alpha_{z}, \alpha_{z})$. The parameter $\alpha_{\mu}$ is chosen such that the variance of the beta distribution is equal to the variance of the normal distribution. This leads to 
\begin{equation}
    \alpha_{x} = \frac{1}{2} \left (\frac{\sqrt{3}}{4}b \right )^2 \frac{H_{xx} }{k_BT}- \frac{1}{2},
    \label{eq:alpha_beta_dist}
\end{equation}
where $H_{xx}$ is the diagonal element of the Hessian matrix. Note that by setting $\alpha_{\mu} = 1$ in the above equation we can find the temperature at which the beta distribution is equivalent to a uniform distribution, this value is 
\begin{equation}
    k_BT_c  = \left ( \frac{\sqrt{3}b}{4} \right )^2 H_{xx}.
\end{equation}

Figure~\ref{fig:plm_comparison} shows a comparison of the distribution of $\delta \mathbf r_i$ corresponding to three normal distributions of varying standard deviation and sampled the beta distributions with $\alpha$ given by eq.~\eqref{eq:alpha_beta_dist}. For the remainder of this paper the perturbed lattice method will be identified by BCC Beta.


\section{Application to the Yukawa One-Component Plasma }
\label{sec:MD_methods}

The Yukawa one component plasma (YOCP) serves as an exemplar system for our investigation due to its well-characterized properties and broad relevance across plasma physics, soft matter, and astrophysical contexts. The YOCP consists of $N$ identical particles with charge $Q$ and mass $m$ interacting via the screened Coulomb (Yukawa) potential:
\begin{equation}
    u(r) = \frac{Q^2}{r}e^{- r / \lambda},
\end{equation}
where $\lambda$ denotes the screening length characteristic of the medium. The thermodynamic state of the system is completely specified by two dimensionless parameters: the screening parameter $\kappa$ and the coupling parameter $\Gamma$, defined as
\begin{equation}
    \kappa = \frac{a_{\rm ws}}{\lambda}, \quad \Gamma = \frac{Q^2}{a_{\rm ws}k_B T},
\end{equation}
where $a_{\rm ws} = (3/4\pi n)^{1/3}$ represents the Wigner-Seitz radius determined by the equilibrium number density $n$ and $T$ is the thermodynamic temperature. Throughout this work, we employ a dimensionless representation where lengths are normalized by $a_{\rm ws}$ and time is expressed in units of the plasma period $\tau_{\omega_p}$
\begin{equation}
    \tau_{\omega_p} = \frac{2\pi}{\omega_p}, \quad \omega_p = \sqrt{\frac{4 \pi Q^2 n}{m}},
\end{equation}
with $\omega_p$ denoting the plasma frequency.

We conducted MD simulations using the open-source software Sarkas \cite{SILVESTRI2022108245} with system size $N = 8192$ particles and fixed screening parameter $\kappa = 2$. Three distinct coupling regimes were investigated ($\Gamma = 2, 20, 200$) for each initialization method described in the preceding section. All simulations employed a Verlet integrator with an integration timestep of $\Delta t = 1.64 \times 10^{-3} \tau_{\omega_p}$. The short-range nature of the Yukawa potential enabled computational efficiency through linked-cell lists with cutoff radius $r_c = 5.7 a_{\rm ws}$, yielding force calculation precision with relative error $\Delta F \sim 10^{-5}$. Phase-space coordinates (positions, velocities, and accelerations) were recorded at intervals of $5\Delta t$ for subsequent analysis.

\subsection{Comparison of initialization methods}
\label{sec:methods_timing}
\begin{figure*}
    \centering
    \includegraphics[width=\linewidth]{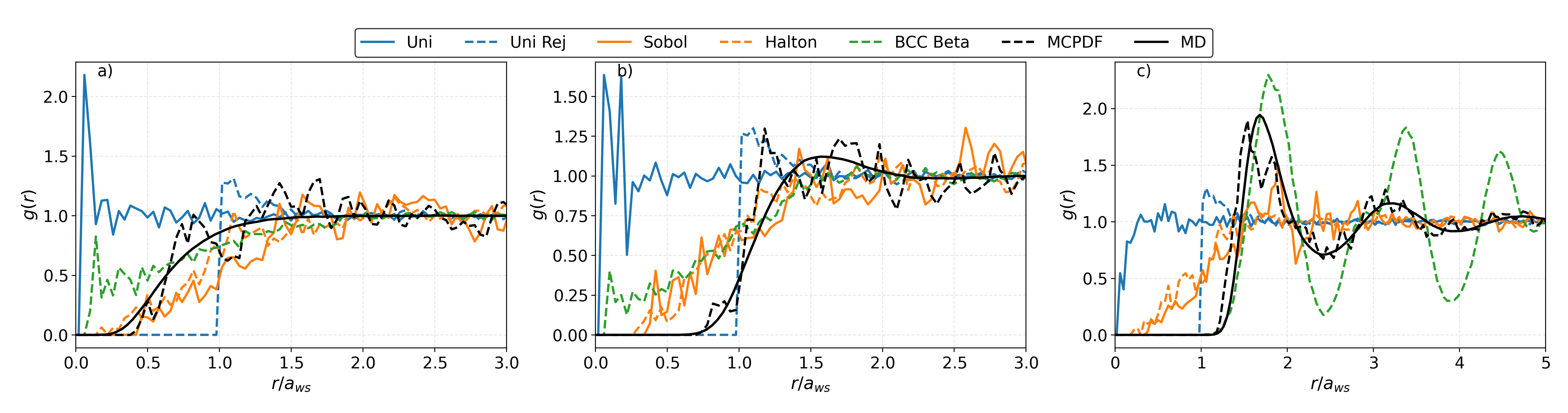}
    \caption{Comparison of radial distribution functions $g(r)$ for different initialization methods for the Yukawa one component plasma for $\kappa=2$, and three coupling parameters: a) $\Gamma = 2$, b) $\Gamma = 20$, and c) $\Gamma = 200$. The radial distance $r$ is normalized by the Wigner-Seitz radius $a_{\rm ws}$. Methods shown are: uniform from Sec.~\ref{subsec:random}, Sobol from Sec.~\ref{subsubsec:Sobol}, Halton from Sec.~\ref{subsubsec:Halton}, Uni Rej from Sec.~\ref{subsec:random_plus_reject}, BCC Uni from Sec.~\ref{subsec:lattice_initialization}, BCC Beta from Sec.~\ref{subsec:BCCBeta}, and MCPDF from Sec.~\ref{subsec:MCPDF}.}
    \label{fig:rdf0_comparison}
\end{figure*}
Here we apply each method of Section~\ref{sec:init_methods} to the YOCP and compare the RDF produced by the methods as well as their computational scaling. Recall that the labels of the methods are summarized in Table~\ref{tab:initialization_methods}.
Figure~\ref{fig:rdf0_comparison} shows a comparison of the RDF, $g(r)$, for the different initialization methods at three coupling strengths ($\Gamma = 2,\, 20, \, 200$), see Section~\ref{sec:MD_methods} for definition of $\Gamma$. The solid black line is the $g(r)$ obtained from averaging the final MD result. The stochastic methods - Uni, Uni Rej, Sobol, and Halton - generate particle configurations independent of thermodynamic parameters, resulting in consistent RDF profiles irrespective of coupling strength. These methods exhibit characteristic features of uncorrelated distributions, with minimal structural organization evident at all coupling parameters.

In contrast, the MCPDF and BCC Beta methodologies demonstrate coupling-dependent behavior that more accurately reproduces the expected particle correlations at each thermodynamic state point. At $\Gamma = 200$, the BCC Beta method manifests pronounced oscillatory patterns in the RDF, with well-defined peaks corresponding to the underlying BCC lattice structure. These oscillations systematically diminish with decreasing coupling strength, approaching a more homogeneous distribution at $\Gamma = 2$. This behavior highlights the method's capacity to effectively transition from crystalline to fluid-like configurations as the coupling parameter decreases.

\begin{figure}
    \centering
    \includegraphics[width=\linewidth]{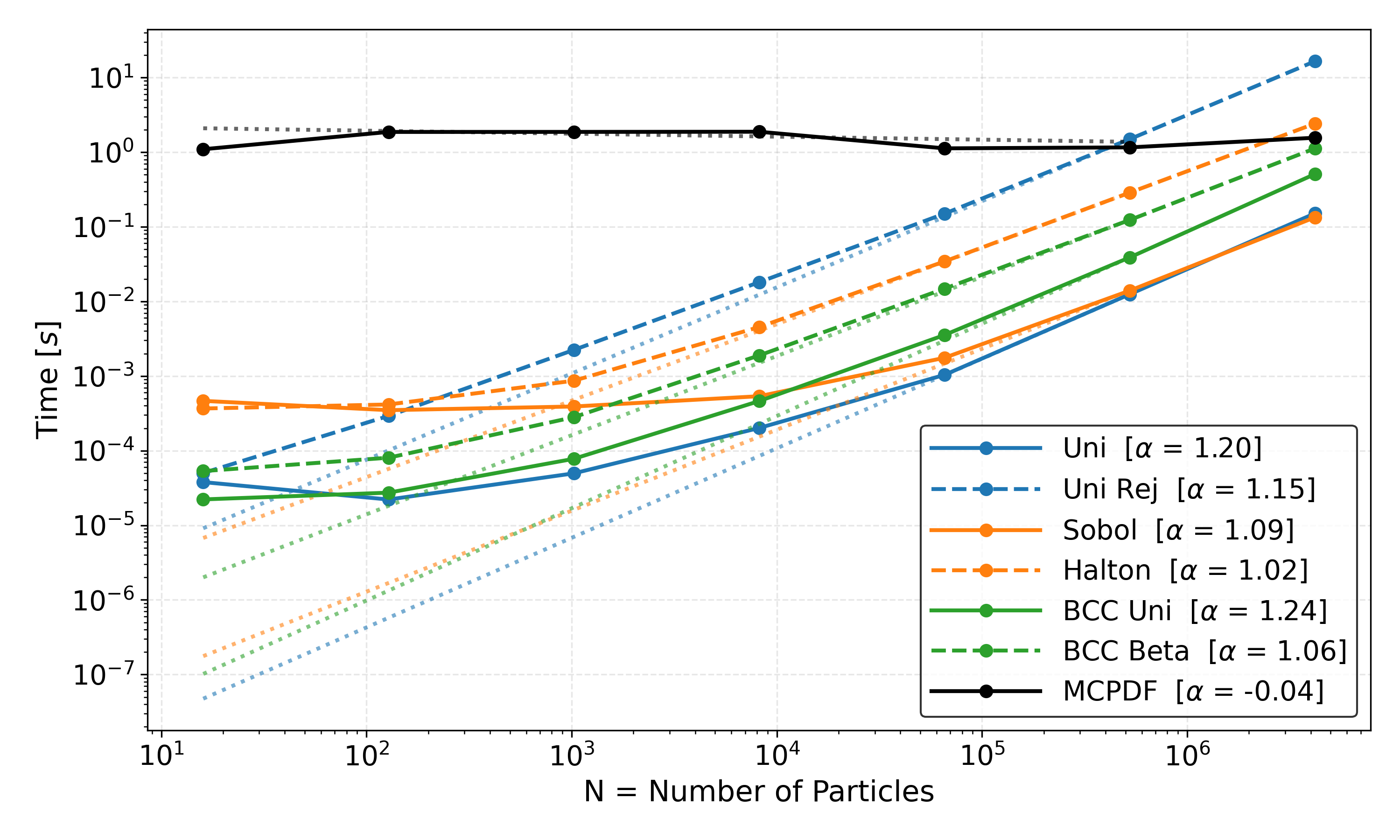}
    \caption{Computational performance comparison of particle initialization methods for the YOCP at $\kappa=2$, $\Gamma=200$. The plot demonstrates scaling behavior by showing initialization time (in seconds) as a function of particle count $N$ for seven distinct initialization approaches: (blue line) uniform correspond to the method in Sec.~\ref{subsec:random}, (red line) Uni Rej to Sec.~\ref{subsec:random_plus_reject}, (orange line) Sobol to Sec.~\ref{subsubsec:Sobol}, (green line) Halton to Sec.~\ref{subsec:qmc_sequences}, (purple line) BCC Uni to Sec.~\ref{subsec:lattice_initialization}, and (brown line) BCC Beta to Sec.~\ref{subsec:BCCBeta}. Dashed lines indicate power law fits ($t \propto N^\alpha$) with scaling exponents $\alpha$ shown for each method. Each data point represents the average of five independent trials, with error bars omitted for clarity.}
    \label{fig:methods_timing}
\end{figure}

Quantitative analysis of computational performance reveals scaling behavior consistent with theoretical expectations. Figure \ref{fig:methods_timing} demonstrates the initialization time as a function of particle count $N$ for all seven methods, with particle numbers ranging from $10^1$ to $10^6$. Power-law regression analysis yields scaling exponents ($t \propto N^\alpha$) for each method, with all algorithms exhibiting near-linear scaling ($\alpha \approx 1$), except for the MCPDF which in this we use the $\mathcal{O}(N^0)$ method detailed at the end of Sec.~\ref{subsec:MCPDF}.

The implementation architecture significantly influences absolute performance metrics. The BCC Uni, BCC Beta, and Uni Rej methods benefit from just-in-time compilation via Numba, while Sobol and Halton leverage optimized SciPy routines. The Uni Rej method, specifically optimized using linked-list data structures, achieves scaling exponent $\alpha = 1.15$, approximating the theoretical linear scaling. 

For small system sizes ($N < 10^3$), all methods except MCPDF demonstrate comparable performance within an order of magnitude, with initialization times dominated by interpreter overhead. As system size increases, the performance differentials become more pronounced, with initialization times spanning four orders of magnitude at $N = 10^6$. The Uni and Sobol methods maintain superior efficiency across all system sizes, while the MCPDF method exhibits substantially higher computational cost.

When considering both structural accuracy and computational efficiency, the BCC Beta method provides an optimal compromise for strongly coupled systems ($\Gamma = 200$), reproducing the expected structural correlations with reasonable computational overhead. For weakly coupled systems ($\Gamma < 20$), the Uniform or Sobol methods offer superior efficiency without significant structural disadvantages.

\subsection{Uncertainty Quantification}
\label{sec:UQ}
The extent to which an MD system is equilibrated depends on the desired properties of interest and is limited by inherent statistical fluctuations and finite-data. Ideally, one would terminate the thermalization cycle when the uncertainty in the desired set of simulation outputs, \textit{e.g.} pressure or thermal conductivity, is computed to within a specified tolerance. \textit{A priori} it is generally impossible to know this when the threshold has been reached, but if an estimate is known for each desired property, one can generate termination conditions consistent with these expectations. 

In this paper, we apply models for the self-diffusion coefficient, the excess internal energy and the viscosity as benchmark functions that inputs the MD temperature deviation and output the expected deviations in each output parameter. Our model for the viscosity is that of the IYVM fit to the Yukawa one component plasma (YOCP) \cite{Johnson2024,MURILLO200849}. The excess energy fit is a combination of the small $\kappa\lesssim1$ fit \cite{10.1063/1.467955}, and the large $\kappa \gtrsim 1$ fit in \cite{PhysRevE.56.4671}\footnote{We believe there is a typo in Eq.(17) of \cite{10.1063/1.467955} where the $b$ coefficients should have the opposite sign. This reduces the deviation from their MD results by an order of magnitude, and makes the coefficient continuous with the results of \cite{PhysRevE.56.4671}.}. The self-diffusion coefficient is given in \cite{10.1063/1.1316084}.  

\begin{figure}
    \centering
    \includegraphics[width=\linewidth]{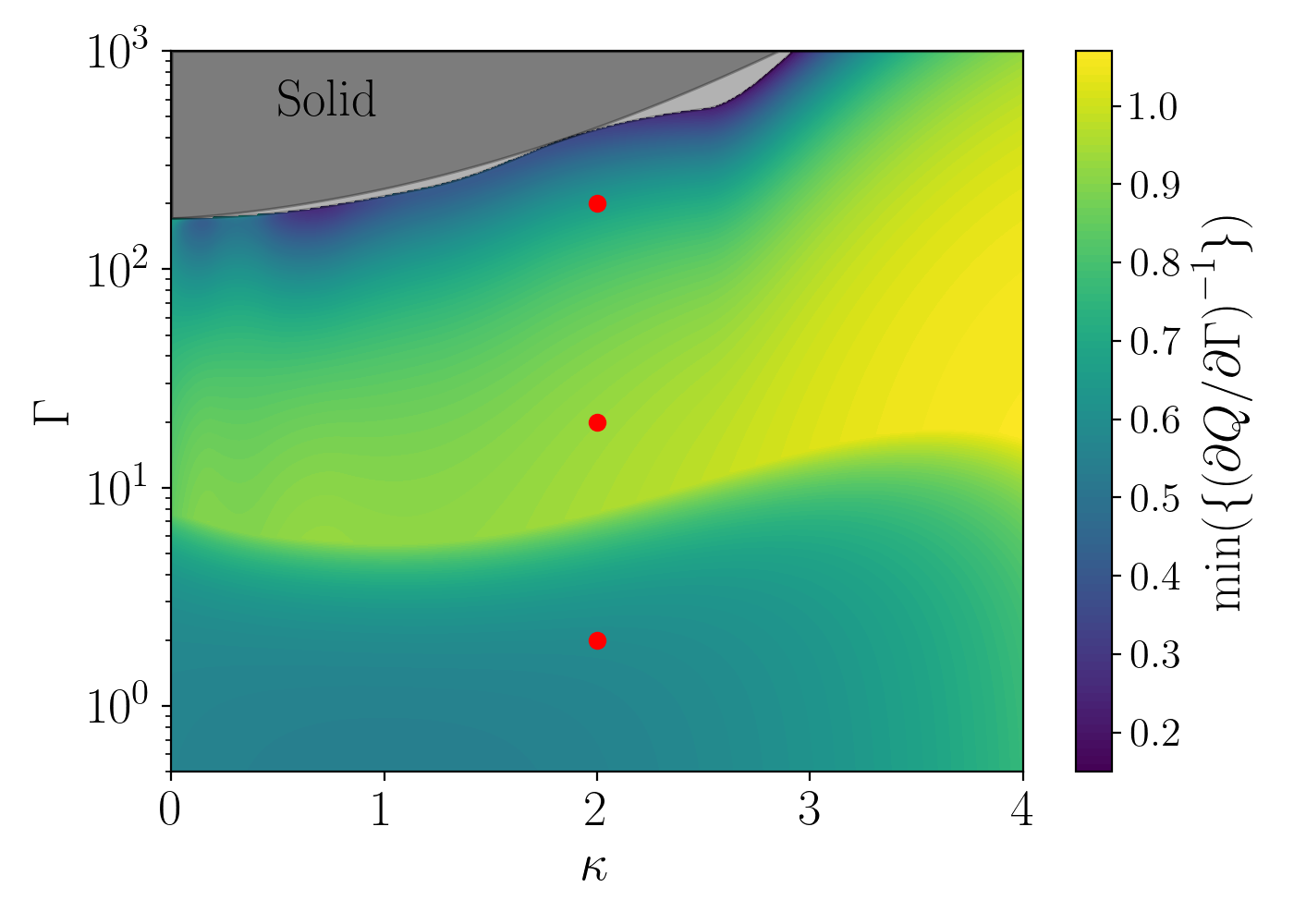}
    \caption{The minimum uncertainty as taken by evaluating Eq.~\eqref{eq:UQ} with $\epsilon=1$. The uncertainty is dominated by viscosity in the lower left corner, which is below the viscosity minimum, and otherwise dominated by the self-diffusion coefficient. The red points correspond to the MD simulations done in this paper at $\kappa=2$, and $\Gamma=(2,20,200)$. There are two lines demarcating the plasma-solid transition corresponding to interpolated data from \cite{10.1063/1.1316084} and a fit from \cite{MURILLO200849}.} 
    \label{fig:UQ_plot}
\end{figure}

Based on these fits, we then compute the minimal temperature deviation expected to yield the desired property to within $1\%$, with, 
\begin{align}
    \frac{\Delta T}{T} \leq \epsilon Min\Big\{ \Big| \frac{d \ln D }{d\ln \Gamma}\Big|, \Big| \frac{d \ln \eta }{d\ln \Gamma}\Big|,  \Big| \frac{d \ln u_{\rm ex} }{d\ln \Gamma}\Big| \Big\},\label{eq:UQ}
\end{align}  
where $\epsilon$ is the desired accuracy of the output quantities, e.g. $\epsilon=0.1$ means we require all three quantities in Eq.~\eqref{eq:UQ} within $10\%$. Note we used the fact that $\Gamma\propto 1/T_i$, and the fact that $\kappa$ is a function primarily of $T_e$, which is a fixed input of the simulation, and thus not sensitive to MD fluctuations with only ions. We plot this quantity assuming $\epsilon=1$ in Fig.~\ref{fig:UQ_plot}, where we can see an estimate that obtaining an error in the three quantities of interest of less than $1\%$, $\epsilon=0.01$, requires a temperature deviation below anywhere from $0.1-1\%$ depending on the point in parameter space. 

An additional difficulty with the determination of thermalization is that inherent statistical fluctuations of the system will result in a physical uncertainty in the temperature itself. The fluctuation in the NVE ensemble of the kinetic energy per particle is\cite{AllenTildesley_Book_2017}
\begin{align}
    \langle \delta K^2 \rangle_{\rm NVE} = \langle \delta K^2 \rangle_{\rm NVT} \left(1 - \frac{C_V^{\rm ideal}}{C_V}\right) 
\end{align}
where $\langle \delta K^2 \rangle_{\rm NVT} = \frac{3}{2} N k_B^2 T^2$, and $C_V$ is the heat capacity which in the ideal limit is $C_V^{\rm ideal}=\frac{3}{2} k_B T$. The corresponding temperature fluctuation is 
\begin{align}
    \frac{\delta T_{\rm NVE}}{T} = \sqrt{ \frac{1}{\frac{3}{2} N} \left(1 - \frac{C_V^{\rm ideal}}{C_V} \right) }.
\end{align}
If one averages the temperature over many uncorrelated time stamps, $M$, one can obtain a better estimate of the mean temperature with error decreasing by an additional factor of $1/\sqrt{M}$. 

For our simulations at $N=8912$ particles, the natural statistical fluctuations corresponds to an expected statistical fluctuation of the temperature of less than $1\%$, which implies a temperature-fluctuation-induced error in our expected transport and EOS properties of $\lesssim 1-10\%$.

\subsection{Thermostat Implementation}
Thermostats drive the system to a desired temperature, loosely mimicking an NVT ensemble. We compared two thermostats: Berendsen and Langevin. It can be shown that in both thermostats the temperature evolution follows the form
\begin{equation}
    T(t) = T(0) e^{-t/ \tau_{B,L}} + \left(1 - e^{-t/\tau_{B,L}} \right)T_d
\end{equation}
where $T_d$ is the desired temperature and $\tau_{B,L}$ represents the coupling strength for the Berendsen ($\tau_B$) or Langevin ($\tau_L$) thermostat. Detailed implementations of both thermostats are provided in Appendix~\ref{app:Thermostats}.

To quantify the influence of thermalization duration on equilibration efficiency, we implemented three distinct canonical ($NVT$) phase durations: $\tau_{_{NVT}} = [1.0, 2.0, 4.0]\tau_{\omega_p}$. The thermostat coupling strength was systematically calibrated to achieve the equilibration criterion $\exp(-t/\tau_{B,L}) = 10^{-2}$, establishing that the system temperature approaches within 1\% of the target value after time $t$, a value motivated by the natural statistical fluctuations in the previous section. Setting $t = 0.5\tau_{_{NVT}}$, we derived coupling parameters for each thermalization period
\begin{equation}
    \tau_{B,L} = - 0.5\tau_{_{NVT}} / \log(0.01),
\end{equation}
corresponding to a strong ($\tau_{_{NVT}} = 1.0 \tau_{\omega_p}$), moderate ($\tau_{_{NVT}}= 2.0 \tau_{\omega_p}$), and weak ($\tau_{_{NVT}} = 4.0 \tau_{\omega_p}$) thermostat.

The simulation protocol comprised 10 sequential, alternating phases of $NVT$ and $NVE$ dynamics. The $NVE$ phase duration was established at five times the corresponding $NVT$ interval to provide sufficient relaxation time for equilibration assessment. Statistical robustness was ensured by executing five independent simulations with distinct pseudorandom number generator seeds for each parameter configuration.

Two contrasting thermostating cycles were implemented to comprehensively evaluate equilibration efficiency: \textit{(i)} ON-OFF cycles initiating with $NVT$ dynamics followed by $NVE$ evolution, and \textit{(ii)} OFF-ON cycles beginning with $NVE$ dynamics subsequently regulated by $NVT$ thermostating.


\section{Results of Molecular Dynamics Simulations}
\label{sec:results}
\begin{figure*}[ht]
    \centering
    \includegraphics[width=\linewidth]{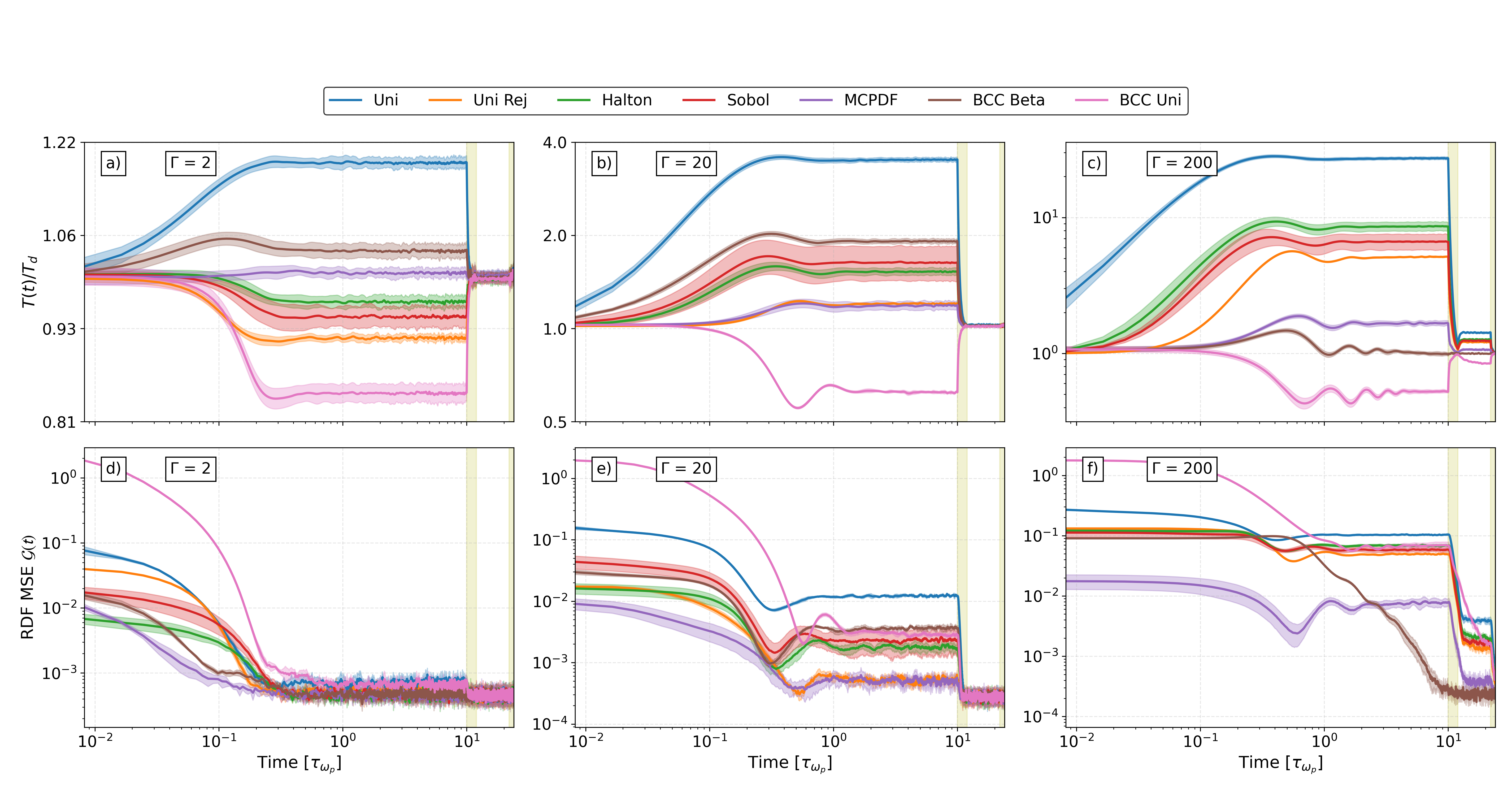}
    \caption{Plots comparing the temperature evolution $T(t)/T_d$ (top panels) and MSE of $g(r)$, see eq.~\eqref{eq:rdf_mse}, (bottom panels) for different initialization methods. The solid line indicates the average over five different runs and the surrounding shaded area indicates the standard deviation. The yellow vertical shaded areas indicate the times at which a Langevin thermostat was turned on. The thermostating cycle used was OFF-ON. The length of the $NVT$ phase is 2.0$\tau_{\omega_p}$ corresponding to a medium strength thermostat. Note that the y-axis in the temperature plots is log scaled, the y-ticks labels have been modified for easier reading.}
    \label{fig:MD_TempMSE_p1.0_Langevin_c01}
\end{figure*}

We begin by examining systems evolving without thermostat intervention, focusing on both the temperature dynamics and structural equilibration across different initialization methods. Figure~\ref{fig:MD_TempMSE_p1.0_Langevin_c01} shows the ratio of instantaneous temperature to target temperature $T(t)/T_d$ (upper panels) and the mean squared error (MSE) of the radial distribution function $g(r)$ (lower panels), defined as 
\begin{equation}
\label{eq:rdf_mse}
    \mathcal G(t) = \int_0^{\infty} dr \left ( g(r, t) - g_{_{NVE}}(r) \right )^2,
\end{equation}
where $g_{_{NVE}}(r)$ is the radial distribution function calculated in the last $NVE$ phase averaged over the $\tau_{_{NVE}}$ time steps. The shaded vertical areas indicate times when the Langevin thermostat was active ($NVT$ phases), with thermostat strength $\tau_{B,L} = -1/\log(0.01)$ corresponding to $\tau_{_{NVT}} = 2 \tau_{\omega_p}$ and $\tau_{_{NVE}} = 10 \tau_{\omega_p}$.  In our analysis, we also calculated the Kullback-Leibler (KL) divergence between the observed velocity distributions and the target Maxwell-Boltzmann distribution to quantify deviations from equilibrium statistics. However, our findings indicate that the KL divergence metric closely tracks the temperature evolution and does not provide significant additional diagnostic information beyond what can be directly inferred from temperature measurements.

The temperature evolution shows two behaviors directly correlated with initialization methodologies~\cite{Murillo2006_DIH,Murillo_DIH_Pop}. At coupling strength $\Gamma = 2$, we observe disorder-induced heating (DIH) in the Uni, MCPDF, and BCC Beta configurations, while order-induced cooling (OIC) manifests in the Sobol, Halton, Uni Rej, and BCC Uni configurations. 

At $\Gamma = 20$ and $\Gamma = 200$, only BCC Uni continues to exhibit OIC behavior, while all other methods display varying degrees of DIH, with BCC Beta showing the least amount of DIH at $\Gamma = 200$. The temperature evolution also exhibits decreased statistical variance with increasing values of $\Gamma$, consistent with the temperature-dependent variance of the initial velocity distributions.

\begin{figure*}[tp]
    \centering
    \includegraphics[width=\linewidth]{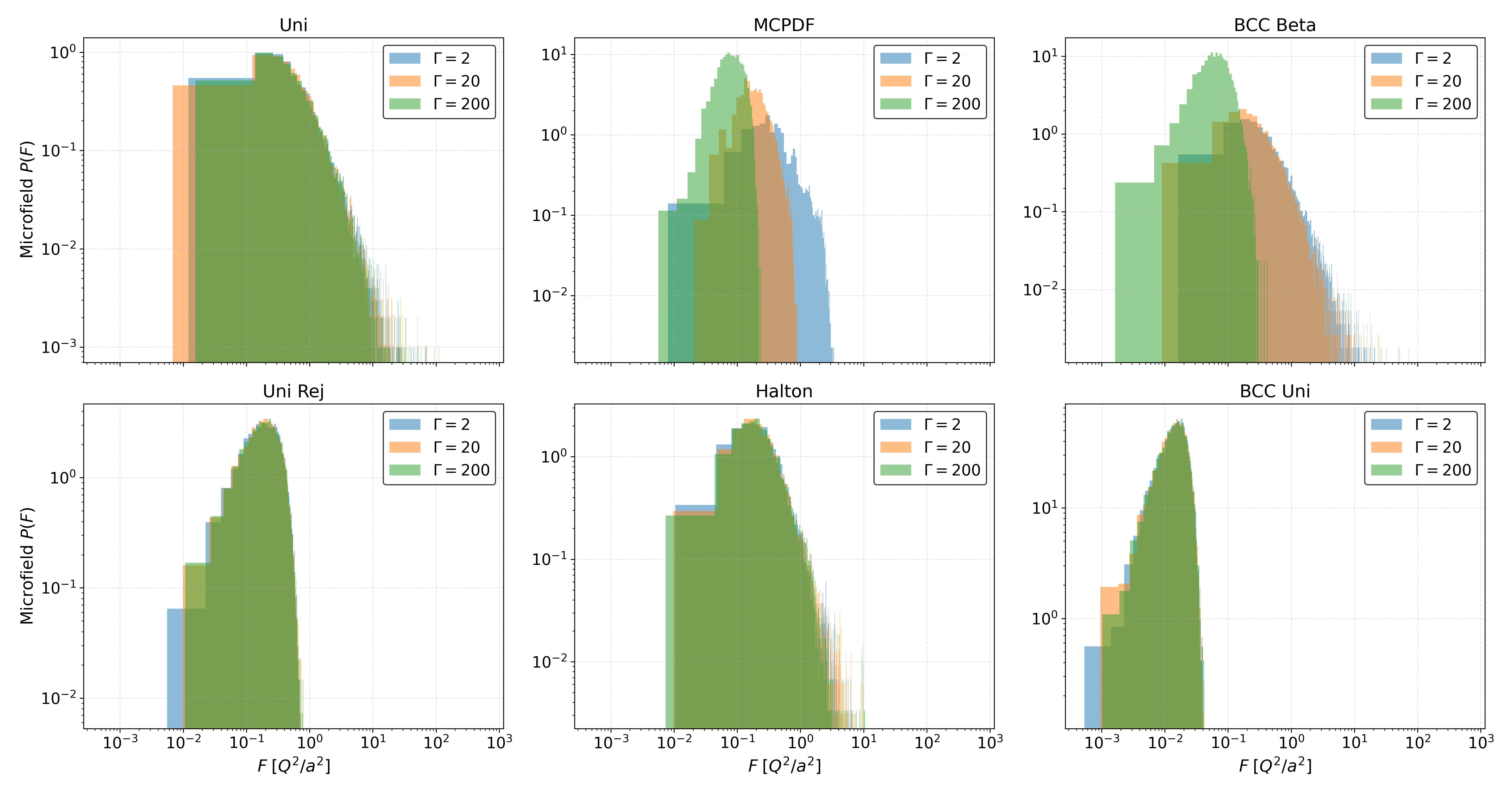}
    \caption{Plots comparing the initial microfield distribution at the three coupling considered for the different initialization methods.}
    \label{fig:microfields}
\end{figure*}
The short-time dynamics of these systems can be comprehensively understood through the lens of microfield distributions $P(F)$ in conjunction with spatial correlation characteristics. Figure~\ref{fig:microfields} shows the initial microfield distributions at three coupling strengths for different initialization methods, revealing differences that directly explain the thermal behaviors observed.

In less correlated configurations like the Uni method (top left panel), while the distribution does have a peak (somewhat obscured by the log-log plotting), it occurs at notably higher force magnitudes with substantially broader spread and significant high-force tail contributions compared to other methods. This distribution pattern indicates that individual particles frequently experience larger forces from nearest neighbors with less effective force cancellation. Consequently, particles respond primarily to these stronger, less balanced forces, manifesting as disorder-induced heating (DIH).

Highly correlated initial configurations like BCC Uni (bottom right panel) exhibit distributions with extraordinarily pronounced peaks at markedly lower force magnitudes. The peak height for BCC Uni is dramatically higher than all other methods—by approximately an order of magnitude—reflecting the exceptional degree of force cancellation achieved in this highly ordered configuration. This extreme force cancellation explains why BCC Uni consistently demonstrates the most significant cooling behavior across all coupling regimes.

The evolution of correlations and their effect on the microfields are highlighted in the MCPDF (top middle) and BCC Beta (top right) distributions which exhibit a clear progression toward sharper, more defined peaks as $\Gamma$ increases from 2 to 200. This progressive sharpening indicates improving force cancellation with increasing coupling strength, consistent with their thermal behavior becoming more ordered-like at higher $\Gamma$ values. The modest peak heights of these methods compared to BCC Uni, however, explain why they still predominantly exhibit heating rather than cooling behaviors, albeit with reduced intensity compared to the Uni method.

The initial MSE values at $t=0$ quantify the structural differences between each initialization method observed in Fig.~\ref{fig:rdf0_comparison}. BCC Uni consistently exhibits the highest initial error across all coupling regimes, followed by the uniform (Uni) method, while the MCPDF method demonstrates the lowest $\mathcal{G}(t=0)$ values. The magnitude of these initial deviations corresponds directly to the temperature excursions observed in the upper panels, where methods with higher initial RDF MSE values typically demonstrate larger temperature fluctuations during equilibration. 

The observed increase in $\mathcal{G}(t = 0)$ with increasing $\Gamma$ is attributable to the increasingly oscillatory behavior of $g(r)$ at higher coupling strengths, characterized by pronounced peaks and troughs at large $r$. Notably, $\mathcal{G}(t = 0)$ for the MCPDF method remains approximately constant at $\sim 10^{-2}$ across all coupling regimes, consistent with its explicit sampling of the system's exact equilibrium $g(r)$. While the BCC Beta method successfully captures the position of peaks and troughs in $g(r)$, it fails to accurately reproduce their magnitude. Interestingly, despite the overly pronounced structure, we can see in Fig.~\ref{fig:MD_TempMSE_p1.0_Langevin_c01} that the BCC beta method arrives nearly precisely on the desired temperature, hinting that a rough characterization as ordered or disordered is likely too simplistic.

The Uni Rej method exhibits a particularly low initial RDF MSE at $\Gamma = 20$, where the rejection radius ($r_{rej} = a_{ws}$) closely corresponds to the effective repulsion sphere between particles. Similarly, the quasi-random sequence methods (Halton and Sobol) generate distributions with $g(r)$ values closer to unity at $\Gamma = 2$, rendering them especially suitable for weakly coupled systems.

At $\Gamma = 2$ and $\Gamma = 20$, the MSE of all methods naturally relaxes to the equilibrium $g(r)$ without thermostat intervention within $t \sim \tau_{\omega_p}$. This efficient self-equilibration results from the relatively high kinetic energy. The higher particle velocities at these coupling strengths facilitate comprehensive exploration of the available phase space, enabling rapid sampling of configurational states and efficient crossing of energy barriers that separate metastable configurations. Furthermore, $g(r)$ at these coupling regimes exhibits only short-range order features with a primary exclusion region, requiring less structural reorganization to achieve equilibrium. The combination of enhanced thermal mobility and simplified target structure allows the system to rapidly converge to equilibrium through natural dynamical processes, circumventing the need for external temperature regulation mechanisms that become essential at higher coupling strengths where thermal motion alone is insufficient to overcome the stronger interparticle potential energy barriers.

At $\Gamma = 200$, only the BCC Beta and MCPDF methods demonstrate sufficiently small values of $\mathcal{G}(t = 10 \tau_{\omega_p})$, indicating that their initial configurations closely approximate the highly ordered equilibrium state characteristic of strongly coupled systems. The minimal perturbations applied to the BCC lattice at high coupling preserve the underlying ordered structure while enabling sufficient local adjustments to approach equilibrium, requiring only minor modifications during simulation progression.

Detailed examination of $g(r)$ at $t = 10 \tau_{\omega_p}$ (not shown here for brevity) reveals that the residual MSE of approximately $10^{-2}$ is primarily attributable to statistical noise between the instantaneous $g(r)$ averaged over 5 time steps and the final $g(r)$ which is averaged over a substantially larger number of time steps, rather than representing genuine structural discrepancies.

As evident in Fig.~\ref{fig:MD_TempMSE_p1.0_Langevin_c01}, all initialization methods successfully achieve the desired temperature $T_d$ after a single thermostat application (yellow shaded area) for weakly and moderately coupled systems ($\Gamma = 2, 20$), while strongly coupled systems ($\Gamma = 200$) require multiple thermostat applications. To quantitatively assess thermostat efficiency, we calculate the average temperature deviation for each $NVE$ phase using
\begin{equation}
    \left \langle \frac{|\Delta T|}{T_d} \right \rangle = \frac{1}{\tau_{_{NVE}}}\int_{0}^{\tau_{_{NVE}}} dt \left | \frac{T(t)}{T_d} - 1 \right |.
    \label{eq:DeltaT}
\end{equation}
This metric allows us to determine the minimum number of thermostat applications required to achieve temperature stability, defined as $\langle \left |\Delta T \right |/T_d \rangle < 0.01$.

Figure~\ref{fig:therm_strength_min_nvt} shows a comparison of thermostat performance at $\Gamma = 200$ across different initialization methods, thermostat types (Berendsen and Langevin), thermostating cycles (OFF-ON and ON-OFF), and thermostat strengths (strong: blue, medium: orange, weak: green). The data reveals several significant trends:

\begin{figure}[hb]
    \centering
    \includegraphics[width=\linewidth]{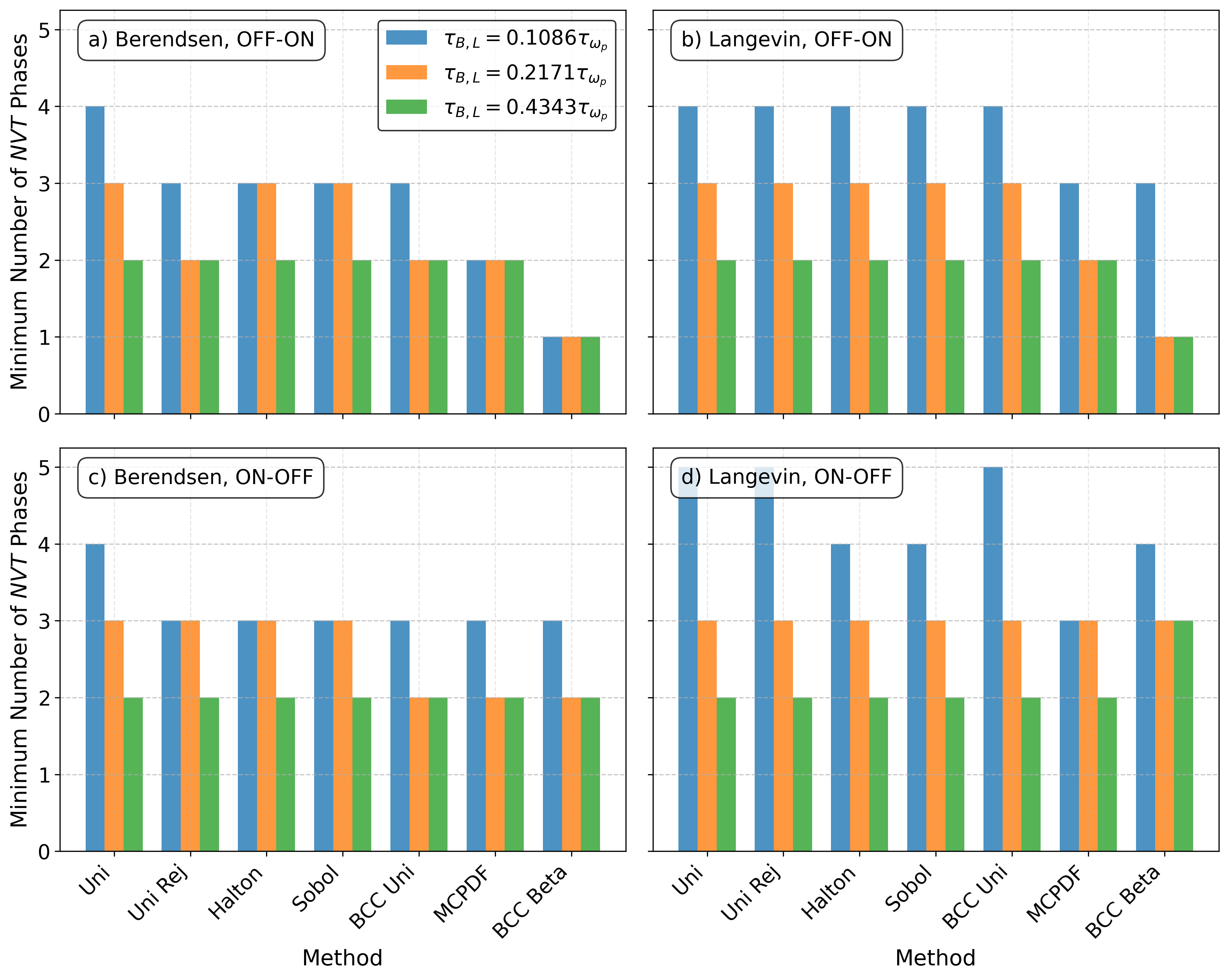}
    \caption{Minimum number of $NVT$ phases required to achieve temperature stability ($\langle |\Delta T| \rangle < 0.01$) at $\Gamma = 200$ for different initialization methods. Results are shown for both Berendsen (left panels) and Langevin (right panels) thermostats with OFF-ON (top panels) and ON-OFF (bottom panels) thermostating cycles. Bar colors represent thermostat strengths: strong ($\tau_{B,L} = -0.5\tau_{\omega_p}/\log(0.01)$, blue, leftmost bars), medium ($\tau_{B,L} = -\tau_{\omega_p}/\log(0.01)$, orange, center bars), and weak ($\tau_{B,L} = -2\tau_{\omega_p}/\log(0.01)$, green, rightmost bars).}
    \label{fig:therm_strength_min_nvt}
\end{figure}

Counterintuitively, weaker thermostat settings (green bars) consistently require fewer $NVT$ phases to achieve temperature stability across nearly all initialization methods, with BCC Beta being the notable exception. This suggests that gentler temperature control allows the system to relax more naturally toward equilibrium. The strong thermostats (blue bars) consistently require the most $NVT$ phases, likely due to their tendency to enforce temperature constraints too rapidly, causing subsequent temperature instabilities when the thermostat is removed.

The BCC Beta method exhibits unique behavior, requiring minimal thermostat intervention regardless of thermostat strength when using the OFF-ON cycle. However, it shows increased sensitivity to thermostat strength in the ON-OFF configuration, requiring 2 and 3 phases for Berendsen and Langevin thermostats, respectively. This distinctive response can be attributed to BCC Beta's highly optimized initial configuration, which closely approximates the equilibrium structure of strongly coupled systems, making it particularly responsive to the timing and sequence of thermostat application.

Comparing thermostat types, the Berendsen thermostat generally requires fewer $NVT$ phases than the Langevin thermostat in the ON-OFF configuration, particularly for ordered initialization methods like BCC Uni. This difference likely stems from the Berendsen thermostat's velocity rescaling approach, which preserves the direction of particle trajectories while adjusting only their magnitudes, resulting in less disruption to emerging structural correlations compared to the stochastic forces introduced by the Langevin thermostat.

A comparison between ON-OFF and OFF-ON thermostating cycles reveals significant implications for simulation efficiency and equilibration strategies. The OFF-ON cycle, where the system initially evolves in an $NVE$ ensemble before thermostat application, generally requires fewer thermostat applications across most initialization methods, particularly with the Berendsen thermostat. This advantage is most pronounced for methods that exhibit substantial initial temperature excursions. The performance of the OFF-ON cycle can be attributed to allowing the system to first undergo natural relaxation processes that resolve the most extreme non-equilibrium features. When applied after this initial relaxation period, thermostats can more effectively fine-tune the temperature without fighting against strong initial dynamics. This approach prevents the thermostat from overcorrecting during the initial high-gradient phase, which often leads to compensatory oscillations that require additional thermostat interventions to resolve, a finding that is more pronounced with strongest thermostat shown in blue in Fig.~\ref{fig:therm_strength_min_nvt}. 


The contrasting performance between these cycles highlights that optimal thermostating strategies should consider the specific initialization method employed. Methods producing significant initial temperature deviations benefit substantially from an OFF-ON approach that allows initial relaxation before temperature control, while methods starting closer to equilibrium show less sensitivity to the thermostating sequence. This finding suggests that simulation protocols should be tailored to the initialization method rather than applying one-size-fits-all thermostating strategies.

\begin{figure*}[htp]
    \centering
    \includegraphics[width=\linewidth]{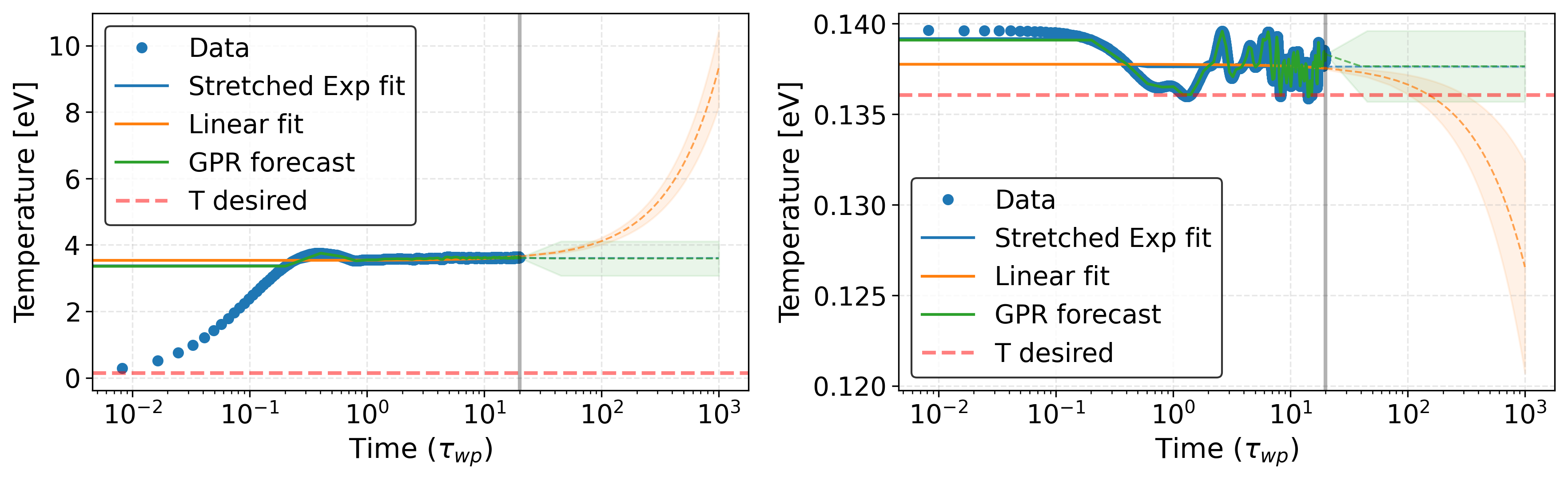}
    \caption{Comparison of temperature evolution forecasting methods for the Uniform initialization with Langevin thermostat with $\tau_{L}=4.0 \tau_{\omega_p}$, and OFF-ON thermostating cycle at $\Gamma = 200$. Left panel: Disorder-induced heating with significant temperature divergence from the target (dashed pink line). Right panel: Thermalized system with temperature oscillations around the equilibrium value. Three models are compared: stretched exponential fit (blue line), linear fit (orange line), and Gaussian Process Regression (GPR, green line). The vertical gray line indicates the boundary between observed data and forecast region. Colored dashed lines represent the forecast of each model and the shaded areas represent prediction uncertainty. Note that the stretched exponential lines are overshadowed by the GPR lines. }
    \label{fig:temp_fits}
\end{figure*}

Equation~\eqref{eq:DeltaT} provides an immediate assessment of temperature stability during brief $NVE$ phases, but a more critical concern is whether this stability persists throughout the entire production phase of the simulation. To address this challenge, we implemented temperature forecasting for the extended production phase by comparing three complementary models: a linear model, a stretched exponential fit, and a Gaussian process regression (GPR) model. Results on two distinct temperature evolution scenarios are shown in Figure~\ref{fig:temp_fits}. The left panel shows a case of disorder-induced heating (DIH) with substantial temperature increase, while the right panel represents a thermalized system exhibiting minor oscillations around the equilibrium temperature. Both scenarios used a Uni initialization method and a weak Langevin thermostat with $\tau_L = 0.4343\tau_{\omega_p}$.

Each forecasting method implements a different mathematical model to capture the temperature dynamics. The linear model represents the simplest approach, using a linear regression of the form $T(t)= T_0+ m t$, where $T_0$ is the initial temperature and $m$ is the slope representing the rate of temperature change. The stretched exponential model captures more complex relaxation processes through the equation 
\begin{equation}
    T(t) = T_{\rm eq} - T_{\rm diff} \exp\left(-\left(\frac{t}{\tau}\right)^\alpha\right),
\end{equation}
where $T_{\rm eq}$ represents the equilibrium temperature, $T_{\rm diff}$ is the temperature difference from equilibrium, $\tau$ is a characteristic relaxation time, and $\alpha$ is the stretching exponent controlling the shape of the relaxation curve. When $\alpha = 1$, the equation reduces to a standard exponential decay, while $\alpha<1$ produces a stretched exponential that has been shown to effectively model relaxation phenomena in complex systems~\cite{Powles2009,LUKICHEV20192983}.

The Gaussian Process Regression (GPR) takes a non-parametric Bayesian approach, modeling the temperature function as a sample from a Gaussian process defined by a mean function (typically zero) and a covariance function (kernel)~\cite{Rasmussen_GPR_Book}. See Appendix~\ref{app:GPR} for details on the kernel we used. 

Each model is fitted to the temperature in the interval $t = [0, 10]\tau_{\omega_p}$ and then used to forecast the temperature to $t = 1000 \tau_{\omega_p}$. The fit is represented with solid lines while the forecast is shown as dashed lines. The uncertainty bands displayed in Figure~\ref{fig:temp_fits} were obtained using parametric bootstrap for the linear and stretched exponential models. For the GPR model, the uncertainty bands represent the posterior predictive distribution, which is a natural output of the Gaussian process formalism~\cite{Rasmussen_GPR_Book}.

As evident in the plots, the linear fit overestimates the temperature at extended time horizons in the DIH scenario, while the stretched exponential and GPR models provide more physically plausible predictions. For the thermalized system (right panel), all three models perform comparably within the observation window and their long-term predictions diverge slightly. The linear fit predicts a future temperature within 10\% of the desired temperature $T_d$, providing a conservative estimate. The stretched exponential and GPR models show more complex behaviors with the GPR notably reverting toward the mean of the training data at extended horizons—a well-known limitation of Gaussian processes when forecasting far beyond the observation range~\cite{Rasmussen_GPR_Book, Roberts_GPR2013}.

The selection of temperature forecasting methodologies was guided by a systematic evaluation of predictive accuracy, computational efficiency, and implementation complexity. While the primary analysis focuses on three models (linear regression, stretched exponential, and Gaussian Process Regression), our investigation encompassed additional time series forecasting techniques including autoregressive (AR) models, however, also these models exhibited mean-reverting behavior at extended time horizons. Furthermore, we investigated the stationarity of the temperature time series using Augmented Dickey-Fuller (ADF) and Kwiatkowski-Phillips-Schmidt-Shin (KPSS) tests. However, quantitative comparison of predictive performance revealed no statistically significant improvement in forecast accuracy with these more complex methodologies.

A critical criterion in model selection was minimization of parameterization complexity imposed on practitioners. Alternative forecasting approaches necessitate additional specification decisions (\textit{e.g.}, AR lag determination, significance thresholds for stationarity tests), potentially introducing user-dependent variability in equilibration protocols. Such variability would undermine the standardization objectives articulated in Section~\ref{sec:intro}.

\section{Conclusions and Outlook}
\label{sec:conclusions}

The systematic investigation of MD equilibration presented in this work establishes a quantitative framework for transforming equilibration from a heuristic procedure to a rigorously quantifiable process with well-defined termination criteria. Through comprehensive evaluation of position initialization methods and thermostating protocols, we have demonstrated that optimal equilibration strategies depend strongly on system coupling strength, with significant implications for computational efficiency in large-scale simulations.

Our findings reveal that at low coupling strengths ($\Gamma = 2$), initialization method selection has minimal impact on equilibration efficiency, with all methods achieving adequate thermalization within comparable timeframes. However, at high coupling strengths ($\Gamma \geq 200$), physics-informed methods such as the perturbed lattice approach (BCC Beta) and the Monte Carlo pair distribution function method (MCPDF) substantially outperform conventional techniques, reducing equilibration time. This performance differential is directly attributable to the initial microfield distributions, where methods generating configurations with balanced force distributions minimize both disorder-induced heating and order-induced cooling effects.

By establishing direct relationships between temperature stability and uncertainties in transport properties, we enable researchers to determine equilibration adequacy based on specified uncertainty tolerances in desired output properties. This approach obviates the need for arbitrary equilibration durations, instead allowing for adaptive termination based on quantifiable prediction of system behavior during the production phase.



The comparative analysis of thermostating protocols revealed that OFF-ON duty cycles, where the system initially evolves in the $NVE$ ensemble before thermostat application, generally outperform conventional ON-OFF approaches, particularly when coupled with weaker thermostat settings. This counterintuitive result suggests that allowing the system to naturally resolve extreme non-equilibrium features before applying temperature control leads to more efficient equilibration pathways.

Based on our comprehensive analysis, we propose the following protocol for optimizing molecular dynamics equilibration across diverse physical regimes: \textit{(i) Identify best initialization method for given system regime:} Determine the appropriate coupling parameter range for your system. For YOCP, this corresponds to the $\Gamma$ value; analogous parameters exist for other pair potentials. In weakly coupled systems, where $g(r)$ is expected to be approximately one, random with reject or quasi-random sequence methods (Sobol, Halton) provide adequate performance with minimal computational overhead. As coupling strength increases one should switch to more physics based methods such as BCC Beta or MCPDF which substantially reduce equilibration time and should be preferred despite their higher computational cost for initialization. \textit{ii) Choose best thermostat cycle:} Both OFF-ON and ON-OFF duty cycles allow for more rigorous determination of equilibration than a single cycle. We advocate an OFF-ON cycle since it allows for a physical configuration to develop before the temperature is changed. We find Berendsen is a rapid way to achieve a thermalized distribution, but physics based model like the Langevin thermostats might be preferred for scientific rigor. \textit{iii) Choose equilibration criterion:} determine output quantities of interest and apply the reasoning of Section~\ref{sec:UQ} to find equilibration termination conditions based on the size of measured temperature fluctuations, as well as the other simple metrics such as radial distribution function convergence.

This protocol transforms MD equilibration from a computationally expensive, trial-and-error process into a systematic procedure with quantifiable termination criteria tailored to the specific physical system and desired output properties. By implementing these approaches, researchers can significantly reduce computational overhead while maintaining or improving simulation accuracy.

Standardization of these equilibration protocols could facilitate more consistent and reproducible results across different research groups, addressing a long-standing challenge in computational physics. We anticipate that the quantitative framework presented here will serve as a foundation for the development of automated equilibration procedures that adapt to the specific requirements of diverse simulation scenarios, further enhancing the efficiency and reliability of molecular dynamics as an investigative tool.
\section*{Acknowledgements}
This work was supported by DOE-DE-SC0025532 and used the MSU Data Machine, which is supported through the NSF Campus CyberInfrastructure program through grant no. 2200792.

\section*{Declaration of interest}
The authors declare that they have no known competing financial interests or personal relationships that could have appeared to influence the work reported in this paper. The research was conducted in the absence of any commercial or proprietary interests. All computational methods described herein were developed and implemented with the sole objective of advancing scientific understanding of molecular dynamics equilibration protocols. The open-source software Sarkas, referenced in this work, is publicly available under standard open-source licensing terms. Any opinions, findings, and conclusions or recommendations expressed in this material are those of the authors and do not necessarily reflect the views of their affiliated institutions or funding agencies.

\bibliographystyle{unsrtnat}
\bibliography{bib}

\appendix

\section{Derivation of Close Pair Probability in Uniform Random Distributions}
\label{appendix:close_pair}

This appendix provides the detailed mathematical derivation of the probability that at least one pair of particles falls within a certain distance in a uniform random distribution. This calculation is relevant for understanding the clumping behavior that occurs in molecular dynamics initialization using the uniform random method described in Section~\ref{subsec:random}.

\subsection{Probability for a Single Pair}

For two particles placed uniformly at random in a cubic box with side length $L$, the probability that they fall within a distance $a$ of each other can be derived from the distribution of their relative coordinates. Let the positions of the two particles be $(x_1, y_1, z_1)$ and $(x_2, y_2, z_2)$, each uniformly distributed in $[0,L]^3$.

The relative displacement between these particles is given by:
\begin{align}
\Delta x &= x_2 - x_1 \\
\Delta y &= y_2 - y_1 \\
\Delta z &= z_2 - z_1.
\end{align}
The difference of two independent uniform random variables on $[0,L]$ follows a triangular distribution on $[-L,L]$ with probability density function
\begin{equation}
f(\Delta) = \frac{L-|\Delta|}{L^2} \quad \text{for } -L \leq \Delta \leq L.
\end{equation}
The joint probability density function of the three differences is:
\begin{equation}
f(\Delta x, \Delta y, \Delta z) = \frac{(L-|\Delta x|)(L-|\Delta y|)(L-|\Delta z|)}{L^6}.
\end{equation}
For $a \ll L$ we can approximate this probability by considering the volume of a sphere with radius $a$ in the space of relative coordinates, weighted by the PDF at the origin:
\begin{equation}
P(d \leq a) \approx \frac{4\pi a^3}{3} \times \frac{1}{L^3} = \frac{4\pi a^3}{3L^3}.
\end{equation}

For a system with $N$ particles, we need to consider all possible pairs. Let $A_{ij}$ denote the event that particles $i$ and $j$ are within distance $a$ of each other. The total number of distinct pairs is $\binom{N}{2} = \frac{N(N-1)}{2}$. To find the probability that at least one pair of particles is within distance $a$, we use the complement approach
\begin{equation}
P(\text{at least one close pair}) = 1 - P(\text{no close pairs}).
\end{equation}
The probability of no close pairs can be written as
\begin{equation}
P(\text{no close pairs}) = P\left(\bigcap_{i<j} \overline{A_{ij}}\right)
\end{equation}
where $\overline{A_{ij}}$ is the event that particles $i$ and $j$ are \textit{not} within distance $a$.

For systems with low particle density, we can approximate these events as independent, yielding:
\begin{equation}
P(\text{no close pairs}) \approx \prod_{i<j} P(\overline{A_{ij}}),
\end{equation}
and for each pair
\begin{equation}
P(\overline{A_{ij}}) = 1 - P(A_{ij}) = 1 - \frac{4\pi a^3}{3L^3}.
\end{equation}
Using $\binom{N}{2} = \frac{N(N-1)}{2}$ total pairs, we get
\begin{equation}
P(\text{no close pairs}) = \left(1 - \frac{4\pi a^3}{3L^3}\right)^{\frac{N(N-1)}{2}},
\end{equation}
which leads to 
\begin{equation}
P(\text{at least one close pair}) = 1 - \left(1 - \frac{4\pi a^3}{3L^3}\right)^{\frac{N(N-1)}{2}}.
\end{equation}

Assuming $\frac{4\pi a^3}{3L^3} \ll 1$, we can use the approximation $(1-x)^n \approx e^{-nx}$ for small $x$:
\begin{equation}
P(\text{at least one close pair}) \approx 1 - e^{-\frac{2\pi a^3 N(N-1)}{3L^3}}.
\end{equation}
This shows that the probability of finding at least one close pair approaches 1 exponentially as $N$ increases. For large systems with many particles, this probability quickly becomes a virtual certainty, explaining why position initialization methods that prevent clumping become necessary as system size grows.

\section{Application to the Yukawa Potential}
The Yukawa potential between two particles at positions $\mathbf{r}_i$ and $\mathbf{r}_j$ in Cartesian coordinates is given by:
\begin{align}
u^Y(r) = \frac{q^2}{r} e^{-\kappa r}
\end{align}
where $r = |\mathbf{r}_i - \mathbf{r}_j|$ is the distance between the two particles.

The gradient of the Yukawa potential with respect to the position vector $\mathbf{r}_i$ is:
\begin{align}
\nabla u^Y(r) = \frac{\partial u^Y(r)}{\partial \mathbf{r}_i} = -\frac{q^2 e^{-\kappa r}}{r^2} \left( 1 + \kappa r \right) \hat{\mathbf{r}}
\end{align}
where $\hat{\mathbf{r}} = \frac{\mathbf{r}_i - \mathbf{r}_j}{|\mathbf{r}_i - \mathbf{r}_j|}$ is the unit vector along the direction of $\mathbf{r}_i - \mathbf{r}_j$.

The Hessian matrix $\vb H_i$ at position $\mathbf{r}_i$ is the matrix of second partial derivatives of the potential:
\begin{align}
H_{i, \mu\nu} = \frac{\partial^2 U(\mathbf{r}_i)}{\partial x_{i\mu} \partial x_{i\nu}} = \sum_{j \neq i} \frac{\partial^2 u^Y(r)}{\partial x_{i\mu} \partial x_{i\nu}}.
\end{align}
The elements of the Hessian matrix of the Yukawa potential are 
\begin{equation}
    H_{\mu\nu} = \frac{q^2e^{-\kappa r}}{r^3}\left ( \frac{x^{\mu}x^{\nu}}{r^2} a(\kappa r) - \delta_{\mu\nu} b(\kappa r) \right ),
\end{equation}
and $a(y) = 1 + y + y^2/3$ and $b(y) = 1 + y$. Given the Hessian matrix in Cartesian coordinates as
\begin{align}
\vb H_i = \begin{pmatrix}
H_{i, xx} & H_{i, xy} & H_{i, xz} \\
H_{i, yx} & H_{i, yy} & H_{i, yz} \\
H_{i, zx} & H_{i, zy} & H_{i, zz}
\end{pmatrix}
\end{align}
we can sample thermal displacements $\mathbf{\delta r}_i$ from a multivariate Gaussian distribution,
\begin{align}
\mathbf{\delta r}_i \sim \mathcal{N}(\mathbf{0}, \vb C_i),
\end{align}
leading to perturbed positions
\begin{align}
\mathbf{r}_i^{\text{perturbed}} = \mathbf{R}_i + \mathbf{\delta r}_i.
\end{align}
In the case of a BCC lattice, each lattice vector has a corresponding lattice vector exactly opposite it, leading to a diagonal Hessian matrix whose elements are 
\begin{align}
    \frac{\partial^2 u^Y(r)}{\partial x_{i\alpha} \partial x_{i\beta}} &= 
    q^2 e^{-\kappa r} \left[ \frac{2 + 2\kappa r + \kappa^2 r^2}{r^3}  \right] \delta_{\alpha \beta},\\
    &= \delta_{\alpha}\delta_\beta {u^Y}''
\end{align}
which allows us to additionally simplify the displacement distribution as the product of three normal distributions. This simplification is true for any cubic lattice with pair potential interactions. By symmetry we only need to consider a single arbitrary lattice point. We compute $H_i = \sum_{j\neq i}^{N_8}{u^Y(r_{ij})}''$, with $N_8$ corresponding to all the neighbors up to the 8th nearest neighbors, once and then sample from the normal distribution $3 N$ times, making the timing of this method comparable to that of one single MD step.





\section{Thermostats}\label{app:Thermostats}
\setcounter{equation}{0}
\renewcommand{\theequation}{C\arabic{equation}}

\subsection{Langevin}

The Langevin thermostat controls the temperature by simulating the interaction of particles with a heat bath through frictional and random forces. The Langevin equation for a particle of mass $m$ and velocity $\mathbf{v}$ is given by:
\begin{equation}
m \frac{d\mathbf{v}}{dt} = \mathbf{F} - \gamma m \mathbf{v} + \mathbf{R}
\end{equation}
where $\mathbf{F}$ represents deterministic forces, $\gamma$ is the friction coefficient, $\mathbf{R}$ is often modeled as Gaussian white noise with a mean of zero and a variance determined by the fluctuation-dissipation theorem.

The fluctuation-dissipation theorem connects the random force's statistics to the temperature $T$ of the system and the friction coefficient $\gamma$, ensuring the system reaches thermal equilibrium. The properties of the $i$th component of the vector $\mathbf{R}(t)$ are:
\begin{equation}
\langle R_i(t) \rangle = 0, \quad \langle R_i(t) R_j(t') \rangle = 2 \gamma k_B T_d m \delta_{ij}\delta(t-t')
\end{equation}
where $T_d$ is the desired temperature, $\delta_{ij}$ is the Kronecker delta, and $\delta(t-t')$ is the Dirac delta function.
This approach not only adjusts the system's temperature but also accurately reproduces the correct physical dynamics under the influence of a heat bath.

Solving the Langevin equation, with the assumption of $F = 0$ for simplicity, leads to 
\begin{equation}
    \mathbf{v}(t) = \mathbf{v}(0)e^{-\gamma t} + \frac{1}{m} \int_0^{t}ds \, e^{-\gamma (t - s)} \mathbf{R}(s).
\end{equation}
Calculating the temperature $T = m \langle v^2(t)\rangle$ we arrive at the temperature evolution equation
\begin{equation}
    T(t) =  T(0) e^{-2\gamma t} +  T_d\left (1 - e^{- 2 \gamma t} \right ).
\end{equation}
The thermostat parameter in the paper is then $\tau_{L} = 1/(2 \gamma)$.

\subsection{Berendsen}
The Berendsen thermostat (BT), unlike the Langevin approach, does not simulate individual particle interactions with a heat bath. Instead, it rescales the velocities of all particles in the system to bring the kinetic temperature towards a target value over a specified relaxation time.
In a strict velocity scaling approach the temperature $T$ is estimated, through a quantity proportional to $\langle v^2 \rangle$, and the velocities are scaled to values consistent with the desired temperature $T_d$, as in $v_i \mapsto \alpha v_i$. In the Berendsen thermostat we begin with a model for the temperature as we would like to see it evolve over a slower
timescale $\tau_{B}$. One model is
\begin{equation}
    \frac{dT}{dt} = \frac{T_d - T}{\tau_{B}}.
\end{equation}
This equation can be solved analytically to yield
\begin{equation}
   T(t) = T(0)e^{-t/\tau_B} + \left(1 - e^{-t/\tau_B}  \right)T_d ,
\end{equation}
which can be seen to transition from the initial temperature $T(0)$ to the desired temperature $T_d$ on a time scale of $\tau_{B}$. 
To implement the thermostat in MD we discretize the Berendsen model across one time step ($\Delta t$) to obtain
\begin{equation}
   T(t + \Delta t) = T(t) + \frac{\Delta t}{\tau_B}\left(T_d - T(t) \right).
\end{equation}
We want to scale the current velocities such that this new temperature $T(t+\Delta t)$ is achieved, because that is the temperature prescribed by the BT. Finding the ratio then of the target temperature and the current temperature, we get
\begin{equation}
   \frac{T(t + \Delta t)}{T(t) } = 1 + \frac{\Delta t}{\tau_{B}}\left(\frac{T_d}{T(t) } - 1 \right).
\end{equation}
Taking the square root of this yields the scaling factor for the velocities:
\begin{equation}
\alpha = \sqrt{ 1+ \frac{\Delta t}{\tau_{B}}\left(\frac{T_d}{T(t) } - 1 \right) }.
\end{equation}

\section{Gaussian Process Regression Forecasting}\label{app:GPR}
\setcounter{equation}{0}
\renewcommand{\theequation}{D\arabic{equation}}
Gaussian process regression requires a kernel describing the assumed correlations between data points. Our implementation employs a composite kernel with 9 hyperparameters
\begin{equation}
    k(t, t') = k_{\rm trend}(t, t') + k_{\rm lt}(t, t') + k_{\rm osc}(t, t'),
\end{equation}
with
\begin{align}
    k_{\rm trend}(t, t') & = \theta_1^2 \left ( 1 + \frac{\left ( t - t'\right )^2}{2 \theta_3\theta_4^2} \right)^{-\theta_5},\\
    k_{\rm lt}(t, t') & =\theta_5^2 \exp \left ( - \frac{(t - t')^2}{2 \theta_6^2} \right ), \\
    k_{\rm osc}(t, t') & = \theta_7^2 \exp \left ( - \frac{(t - t')^2}{2 \theta_8^2}  - 2 \frac{\sin^2( \pi (t - t'))}{\theta_9^2} \right ).
\end{align}
The trend component $k_{\rm trend}(t, t')$ captures medium-scale behavior, the long-term component $k_{\rm lt}(t, t')$ models larger temporal scales, and the oscillatory component $k_{\rm osc}(t, t')$ accounts for temperature fluctuations.

\end{document}